\ifx\ieee\undefined
\documentclass{ifacconf}

\usepackage{amsmath,amssymb,amsfonts,mathtools} 
\usepackage{savesym}
\savesymbol{theoremstyle}
\usepackage{amsthm} 
\restoresymbol{IFAC}{theoremstyle}
\usepackage{dsfont,eucal,bbm,bm,nicefrac} 
\usepackage{graphicx,float,subcaption,booktabs} 
	\graphicspath{{./figures/}}
\usepackage{algorithm} 
\usepackage{algcompatible}
\usepackage{url}            
\usepackage{tikz,xcolor,soul} 
\usepackage{textcomp}
\usepackage{balance}
\usepackage{enumerate}
\usepackage{natbib}   

\else
\documentclass[journal]{IEEEtran}

\usepackage{amsmath,amssymb,amsthm,amsfonts,mathtools} 
\usepackage{dsfont,eucal,bbm,bm,nicefrac} 
\usepackage{graphicx,float,subcaption,booktabs} 
	\graphicspath{{./figures/}}
\usepackage{algorithm,algorithmic} 
\usepackage[hidelinks]{hyperref} 
\usepackage{tikz,xcolor,soul} 
\usepackage{cite}
\usepackage{textcomp}
\usepackage{balance}
\usepackage{enumerate}

\fi

\newtheorem{theorem}{Theorem}

\newtheorem{lemma}[theorem]{Lemma}

\newtheorem{definition}{Definition}
\newtheorem{problem}{Problem}

\newcommand{\pbra}[1]{\ensuremath{\left( #1\right)}}
\newcommand{\sbra}[1]{\ensuremath{\left[ #1\right]}}
\newcommand{\cbra}[1]{\ensuremath{\left\{ #1\right\}}}

\newcommand{\pder}[2]{\ensuremath{\frac{\partial #1}{\partial #2}}}
\newcommand{\E}[1]{\ensuremath{\mathbb{E}\left[ #1\right]}}

\DeclareMathOperator*{\argmin}{arg\,min}

\DeclareMathOperator*{\minimize}{minimize}

\begin{document}

\ifx\ieee\undefined

\begin{frontmatter}

\title{Learning a Network Digital Twin as a Hybrid System\thanksref{footnoteinfo}} 

\thanks[footnoteinfo]{Research partially supported 
by the Swedish Foundation for Strategic Research (SSF) grant IPD23-0019.}

\author[First]{Christos Mavridis} 
\author[Second]{Fernando S. Barbosa} 
\author[Second]{Hamed Farhadi}
\author[First]{Karl H. Johansson}

\address[First]{Division of Decision and Control Systems, 
School of Electrical Engineering and Computer Science,
KTH Royal Institute of Technology, Stockholm, Sweden (e-mails: \{mavridis,kallej\}@kth.se).}
\address[Second]{Ericsson Research, Stockholm, Sweden (e-mails: \{fernando.dos.santos.barbosa,hamed.farhadi\}@ericsson.com).}

\else

\title{
Learning a Network Digital Twin \\as a Hybrid System}

\author{Christos N. Mavridis$^*$, 
Fernando S. Barbosa$^\dagger$, 
Hamed Farhadi$^\dagger$, 
and Karl Henrik Johansson$^*$  
%
\thanks{$^*$%
Division of Decision and Control Systems, 
School of Electrical Engineering and Computer Science,
KTH Royal Institute of Technology, Stockholm.
{\tt\small emails:\{mavridis,kallej\}@kth.se}.}%
\thanks{$^\dagger$%
Ericsson Research, Stockholm, Sweden
\tt\small emails: fernando.dos.santos.barbosa@ericsson.com, 
hamed.farhadi@ericsson.com}.%
\thanks{%
Research partially supported 
by the Swedish Foundation for Strategic Research (SSF) grant IPD23-0019.%
}%
}

\maketitle
 \thispagestyle{empty}
\pagestyle{empty}

\fi

\begin{abstract}
Network digital twin (NDT) models are virtual models that replicate the behavior of physical communication networks and are considered a key technology component to enable novel features and capabilities in future 6G networks. In this work, we focus on NDTs that model the communication quality properties of a multi-cell, dynamically changing wireless network over a workspace populated with multiple moving users. We propose an NDT modeled as a hybrid system, where each mode corresponds to a different base station and comprises sub-modes that correspond to areas of the workspace with similar network characteristics. The proposed hybrid NDT is identified and continuously improved through an annealing optimization-based learning algorithm, driven by online data measurements collected by the users. The advantages of the proposed hybrid NDT are studied with respect to memory and computational efficiency, data consumption, and the ability to timely adapt to network changes. Finally, we validate the proposed methodology on real experimental data collected from a two-cell 5G testbed.
\end{abstract}

\ifx\ieee\undefined


\end{frontmatter}

\else


\fi

\section{Introduction }
\label{Sec:Introduction}

Network digital twins (NDTs) are virtual models that replicate the behavior and structure of physical communication networks, facilitating real-time visualization, simulation, and performance analysis and optimization \citep{NDT_outlook2012}. 
These are treated as one of the key technologies towards the 6G vision of connecting the physical and digital worlds \citep{HexaX6Gvision}. 
These digital counterparts allow network administrators to evaluate operational metrics, foresee potential issues, and experiment with multiple configurations in a risk-free environment, thereby minimizing service disruptions \citep{DTexperimental2025}. 
This approach underpins key advancements such as automated network management, predictive upkeep, and efficient resource deployment, contributing to improved robustness and adaptability of communication networks \citep{6GNDT}. 
%
Given the increasing complexity of communication networks and the advancements in computational hardware enabling the learning of complex systems \citep{QCDT24}, NDTs are anticipated to become essential tools for managing the intricate dynamics of modern communication infrastructures and ensuring the maintenance of high-quality service levels. Consequently, NDTs are expected to play a critical role in future 6G networks \citep{DTsurveyFuture2025}.

In this work, we focus on NDTs that model the communication quality properties of a multi-cell, dynamically changing wireless network over a workspace populated with multiple, moving user equipments (UEs).
These types of NDTs have wide range of applications, including 
resource allocation \citep{al2004routing} and 
communication-aware motion planning,
where the objective is to design optimal decisions for the UEs while maintaining and optimizing the connectivity to a base station
\citep{muralidharan2021communication,licea2019communication}. 
Regarding the network properties, most of the works in the literature 
only considered the signal strength of each UE, ignoring terms related to the
Quality of Service (QoS), such as throughput, end-to-end latency affected by routing, 
or the impact of the UE handover in the presence of multiple base stations 
\citep{muralidharan2021communication,licea2019communication,yan2014go,lindhe2013exploiting}. 
In particular, considering scenarios such as high-band transmissions in densely populated areas with multiple based stations, handover modeling becomes a critical component of a wireless network that need to be well considered in the NDT design and operation.

In addition to the limitation in the representation of network properties, full knowledge of the wireless channel spatial variations is a common assumption in the literature \citep{licea2019communication,lindhe2013exploiting,lindhe2010adaptive}. 
In the absence of this assumption, data-driven methods can be employed to identify the NDT
\citep{gupta2021machine,ghaffarkhah2011communication}.
Most works make use of a probability density model to estimate the network properties 
\citep{yan2014go,ghaffarkhah2011communication,yan2013co}, while others use
model-based learning methods \citep{parasuraman2023rapid}.
However, accurately reconstructing a network model completely from data can be challenging, and sensory measurements can be slow and expensive to acquire
\citep{fink2013robust,gupta2021machine, penumarthi2017multirobot,caccamo2017rcamp,ali2018motion}.
%
%
%
Thus, designing a data-based, explainable, resource-efficient NDT model with fast  adaptation mechanisms still remains an open problem with important implications.
%

\subsection{Contribution}

Towards this direction, we propose the modeling and identification of an NDT model as a hybrid system that
(a) has multiple modes, each corresponding to one of potentially many base stations of the network, and (b) is identified and continuously updated through a data-driven and explainable prototype-based learning algorithm based on principles from hotomopy optimization \citep{mavridis2023multi,mavridis2025real}. 

To the best of our knowledge, the advantages of modeling an NDT as a hybrid system with different operating modes have not been studied before. 
In this work, we show that a hybrid NDT can facilitate not only a more accurate representation of the wireless network, but also the ability to adapt to rapid network changes faster and more efficiently in terms of computation and data usage. 

We develop a hybrid identification algorithm as an extension of a 
prototype-based learning method trained using principles from online deterministic annealing (ODA),
a homotopy optimization approach, first introduced for clustering problems
\citep{mavridis2023annealing,mavridis2025real}. 
ODA makes use of entropy regularization to progressively estimate the number of clusters needed and provide real-time control of the performance-complexity trade-off.
Based on this approach, we construct a combined clustering, classification, and regression algorithm for the identification of an NDT as a hybrid system.

Finally, we compare our methodology against a standard identification algorithm 
in a 5G network with two base stations and event-triggered network changes located at Ericsson’s headquarters, Stockholm, Sweden.


\subsection{Notation}

The sets $\mathbb R$ and $\mathbb R_+$
represent the sets of real and non-negative real numbers, respectively.
Unless otherwise specified, random variables $\mathcal{X}:\Omega\rightarrow \mathbb R^d$ are defined in 
a probability space $(\Omega,\mathbb F,\mathbb P)$.
The probability of an event is denoted 
$\mathbb P\sbra{\mathcal X\in S} \coloneq \mathbb P\sbra{\omega\in\Omega: \mathcal X(\omega)\in S}$, 
and the expectation operator $\E{\mathcal X} = \int_\Omega \mathcal X \textrm{d}\mathbb P$.
In case of multiple random variables $(\mathcal X,\mathcal Y)$ and a deterministic function $f$, 
the expectation operator $\E{f(\mathcal X,\mathcal Y)}$ 
is understood with respect to the joint probability measure, while 
$\E{\mathcal X|\mathcal Y} \coloneq \E{\mathcal X|\sigma(\mathcal Y)}$ denotes 
the expectation of $\mathcal X$ conditioned to the $\sigma$-field of $\mathcal Y$.
Stochastic processes $\cbra{\mathcal X(k)}_k$, $k\in\mathbb Z_+$, are defined in the filtered probability
space $(\Omega,\mathbb F, \cbra{\mathcal F_n}_n,\mathbb P)$, where $\mathcal F_n = \sigma(\mathcal X(k)|k\leq n)$, $k\in\mathbb Z_+$, is the natural filtration. 
The indicator function of the event $\sbra{\mathcal X\in S}$ is denoted $\mathds{1}_{\sbra{\mathcal X\in S}}$. 
%


\section{Hybrid Network Digital Twin Model}
\label{Sec:HybridNDT}

In this work, we are interested in NDT representations that can model both continuous-valued network properties, such as signal strength and latency, and discrete-valued properties, such as cellular association between base stations. 
In this section, we define the properties of a general NDT representation 
and develop an NDT with hybrid structure as a modeling design that enables modular representation and fast adaptation.

\subsection{Network  Properties}
\label{sSec:NetworkProperties}

\begin{figure}[ht]
\centering
\begin{subfigure}[b]{0.24\textwidth}
\centering
\includegraphics[trim=0 0 0 0,clip,width=1.0\textwidth]{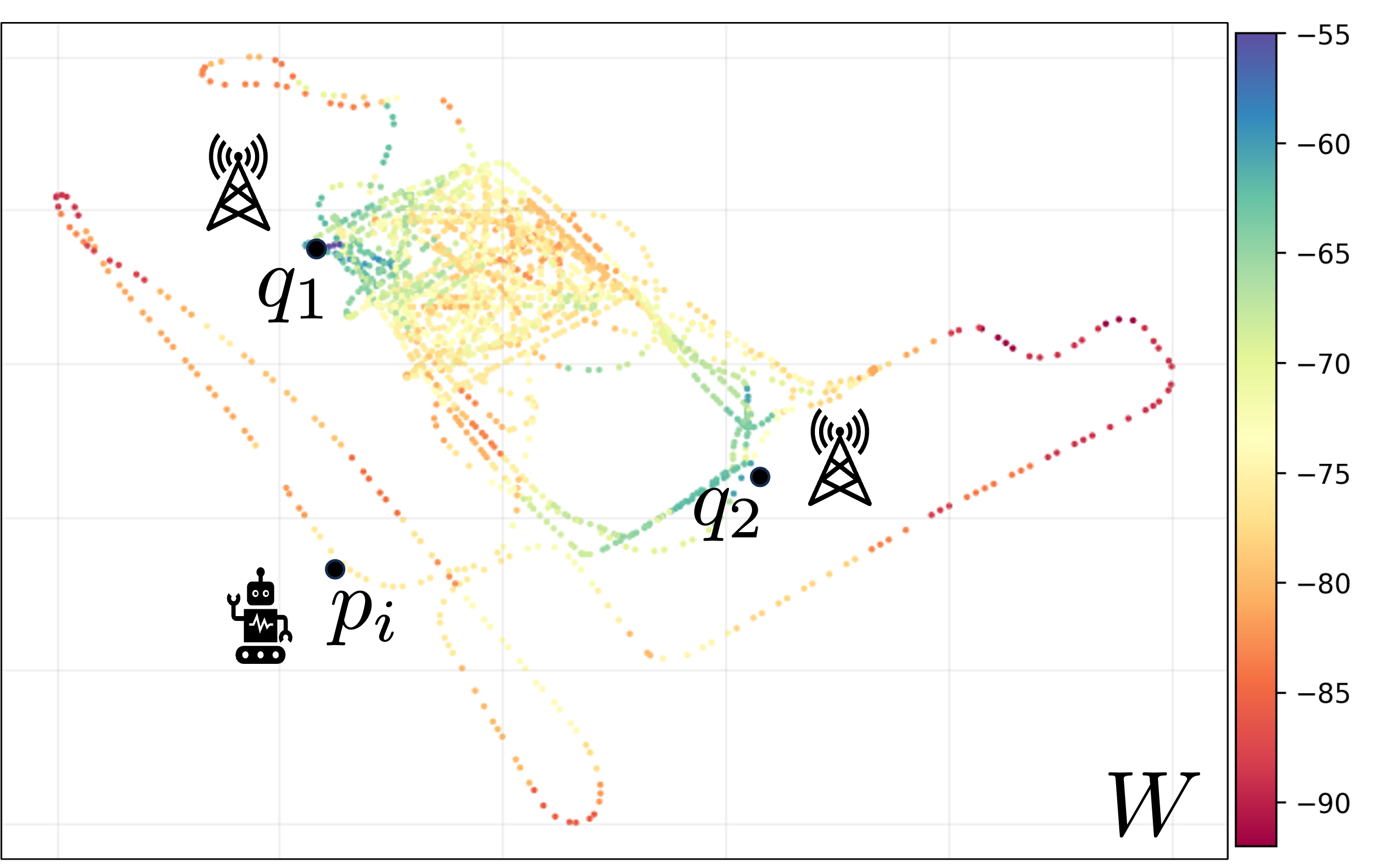}
\caption{User RSRP measurements.}
\end{subfigure}
\begin{subfigure}[b]{0.23\textwidth}
\centering
\includegraphics[trim=0 20 0 0,clip,width=1.0\textwidth]{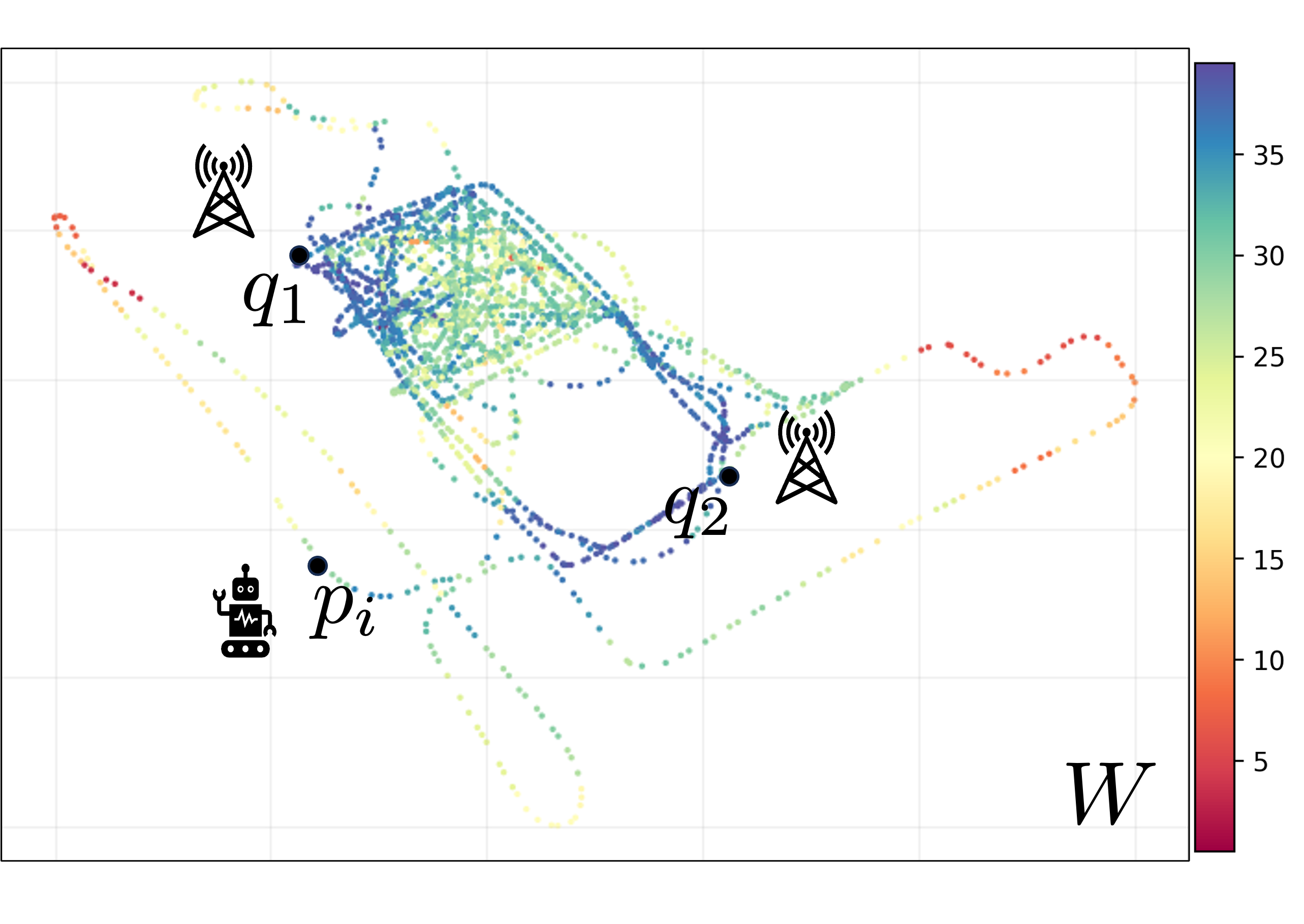}
\caption{User SINR measurements.}
\end{subfigure}
\caption{Illustration of a 2D workspace with two cells and one moving UE providing network measurements.}
\label{fig:input}
\end{figure}

Consider a workspace $W\in \mathbb R^d$ containing $n$ moving agents at locations $p_i(t)\in W$, $i\in \{1,\ldots,n\}$, $t\in\mathbb R_+$, and $m$ base stations 
$\cbra{I-1,\ldots,I_m}$, at fixed positions $q_i\in W,i\in \{1,\ldots,m\}$. 
The workspace $W$ usually represents a two-dimensional ($n=2$) or three-dimensional ($n=3$) space but can be generalized to more dimensions.
A representation of a 2D workspace is depicted in Figure \ref{fig:input}.

%

%
Within the workspace $W$, electromagnetic signals are transmitted carrying information communicated through the network. 
The network is responsible for immediate and fast connectivity and maximization of the communication quality of the users. 
Among the various communication quality metrics that can be measured in a wireless 5G/6G network, 
the most important in communication-aware cyber-physical systems and IoT applications are signal strength, user latency, channel throughput, and communication delay due to handovers between cells \citep{chowdhury20206g}.

\subsubsection{Signal Strength:}

Signal strength comprises mainly three components:
path loss (low-frequency decay in the channel power with distance), 
shadowing (high-frequency fading mainly due to obstructing structures), and multi-path fading (very high-frequency attenuation due to scattering and reflection), given the locations of the base stations 
\citep{muralidharan2021communication,licea2019communication}. 
Signal strength can be measured by numerous metrics, including 
the Reference Signal Received Power (RSRP) and  
the Signal-to-Interference-plus-Noise Ratio (SINR) observed by the user
\citep{dahlman20205g}.

\subsubsection{Latency and Channel Throughput:}

Throughput represents the actual rate at which user data is successfully transmitted over a wireless channel, and depends on SINR, allocated bandwidth, and the radio access technology (e.g. code rate, modulation, and waveform).
Latency is time delay experienced for a data packet to travel from the source to the destination across the network, including transmission delay, processing delay, propagation delay, and queuing delay. Latency is also affected by SINR, network congestion, and radio access technology (e.g. coding, modulation, beamforming,...) \citep{dahlman20205g}. 

%

\subsubsection{Handover Delays:}
A handover is the process of transferring an active connection (voice or data) from one base station to another to maintain service continuity as the user moves or radio conditions change. A handover is triggered when the UE detects that the signal quality (e.g., RSRP or SINR) from a neighboring cell becomes better than that of the serving cell over a time window, or when network conditions require transferring the connection to maintain service quality, or based on the predictions of the signal quality \citep{AIMLEricsson23}. Handovers can increase packet latency and delay due to the time required for signaling, context transfer, and resource reallocation between cells during the connection switch.

%


\subsection{Network  Digital Twin}
\label{sSec:NDT}

Given the wireless network properties of interest in a workspace $W$, 
we define a Network Digital Twin model as follows: 

\medskip
\begin{definition}[Network Digital Twin]
Let $t\in R_+$ represent time, and 
$x\in W\subset \mathbb R^d$ be a position vector in the workspace.
Define $Q(t,x)\in \mathbb R^l$ a vector function of $l$ scalar quantities measuring the communication quality (such as signal strength, latency, or throughput), and 
$c(t,x)\in \{1,2,\ldots,m\}$ a discrete-valued function representing the index of the ID of the cell connected to an agent at position $x$, such that $I=\{I_1,I_2,\ldots,I_m \}$.
Then, a Network Digital Twin (NDT) model $\tilde N$ is defined as a function $N:\mathbb R^{d+1}\rightarrow \mathbb R^{l+1}$, that maps the time and position vector to a vector of communication quality and cell ID index, i.e.,
\begin{equation}
    \tilde N:(t,x) \mapsto (Q,c).
    \label{eq:NDT}
\end{equation}
\label{def:NDT}
\end{definition}

To validate or identify an NDT model of the form \eqref{eq:NDT}, observations $O_i$ can be collected from each agent $i\in \{1,2,\ldots,n\}$, such that:
\begin{equation}
O_i(t,x) = \begin{bmatrix} 
    t \\ 
    x(t) \\
    Q(t, x) \\
    c(t, x)
    \end{bmatrix},
\label{eq:Obserations}
\end{equation}
where, the components of 
$Q(t,x)\in \mathbb R^l$ are quantified by easily obtainable network measurements, including user RSRP, user SINR, communication latency or throughput, etc.
Thus, the observation vector consists of easily obtained measurements and gives sufficient information that best represents the behavior of the network with respect to model \eqref{eq:NDT}.

\subsection{Network  Digital Twin as a Hybrid System}
\label{sSec:HybridNDT}

In this section, we propose the representation of the NDT in \eqref{eq:NDT} 
as a hybrid system. 
The hybrid structure offers a modular and explainable representation of the network properties, simplifies the learning models used in data-driven approaches, and allows quick adaptation to rapid changes that occur to a base station, such as cell malfunction or overload.

A general input-output representation of a hybrid system can be written in the form: 
\begin{equation}
\begin{cases}
    y(t) &= N_{\sigma}(t,x)\\
    \sigma(t) &= \phi(t,x)
\end{cases},\quad t\in \mathbb R_+,
    \label{eq:hybrid}
\end{equation}
where 
$x \in W\subset \mathbb R^d$ represents the input vector,
$y(t) \in \mathbb R^{l}$ represents the output vector,
and $\sigma(t)\in\cbra{1,\ldots, s}$ 
is called the mode-switching signal and can take a finite number 
of $s$ values that are called the modes of the hybrid system. The functions 
$N_i: \mathbb R^{d+1} \rightarrow \mathbb R^{l}$, $\forall i\in \cbra{1,\ldots, s}$,
and $\phi: \mathbb R^{d+1} \times \mathbb{R}^{d} \rightarrow \cbra{1,\ldots, s}$ define the continuous and discrete dynamics of the hybrid system, respectively.

To capture cell-specific properties and handover phenomena, we focus on a special case of hybrid systems, namely switched systems for which $\phi$ is given by:
\begin{equation}
\phi(t,\sigma,x) = c(t,x) ,\quad t\in \mathbb R_+,
\label{eq:phi1}
\end{equation}
such that the NDT has $m$ modes, each one describing the network properties when connected with each base station with ID $I_i$ and position $q_i$, $i=1,\ldots,m$.
This modularity provides the necessary information to detect and predict handovers, as well as a framework to detect and update the NDT model when network changes depend on cell-specific network events, such as cell malfunction or overload.

While \eqref{eq:hybrid}, \eqref{eq:phi1} provide an intuitive hybrid structure for the NDT, 
in practice, the rule $\sigma(t) = c(t,x)$ defines a time-dependent mode switching behavior that is difficult to 
compute, represent with a model, or estimate from observations. 
%
Therefore, we propose an alternative representation. 
We start with a partition-based hybrid representation of the form:
\begin{equation}
\sigma(t) = i \Leftrightarrow x\in S_i(t) ,\quad t\in \mathbb R_+,
\label{eq:phi2}
\end{equation}
where the set 
$\cbra{S_i(t)}_{i=1}^s$ defines a partition
of $W$, i.e., 
$S_i(t)\cap S_j(t) = \emptyset$ for $i\neq j$, and $\bigcup_i S_i(t) = W$, $\forall t\in \mathbb R_+$.
It is easy to see that the representation \eqref{eq:phi2} is equivalent to \eqref{eq:phi1} for a time-dependent partition $\cbra{S_i(t)}_{i=1}^s$ such that $x\in S_i(t) \Leftrightarrow c(t,x)=i$, $\forall t\in \mathbb R_+$. 

\begin{figure}[t]
\centering
\includegraphics[trim=0 20 0 30,clip,width=0.40\textwidth]{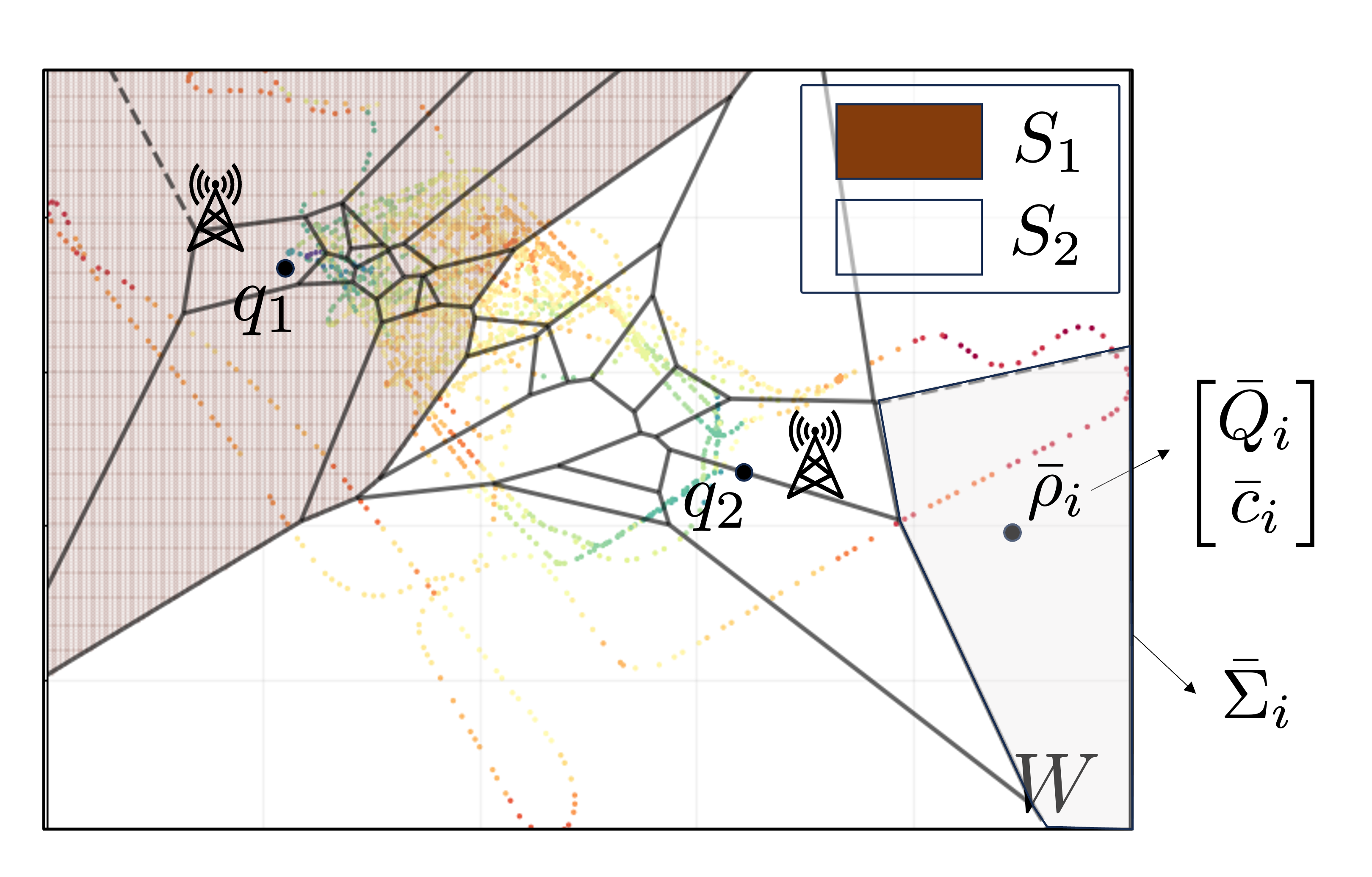}
\caption{Prototype-based representation of the Hybrid NDT.}
\label{fig:hybridNDT}
\end{figure}

To simplify the modeling and learning complexions of this representation, we need to (a) address the time dependency, and (b) find a simple representation of the non-convex regions $S_i(t)$, $i=1,\ldots,s$.
Time dependency will be discussed in Sections \ref{sSec:learning} and \ref{Sec:TimeDependency}.
Regarding the representation of the modes, we introduce a finer polyhedral partition 
$\cbra{\Sigma_i(t)}_{i=1}^K$ of $W$, for some $K>s$, such that 
\begin{equation}
S_i(t) = \bigcup_{j\in J_i} \Sigma_j(t),\quad i=1,\ldots,s,\quad t\in \mathbb R_+,
\label{eq:Sigmas}
\end{equation}
with $J_i\cap J_j= \emptyset$ and $\bigcup_i J_i=K$.
In other words, each $S_i(t)$ can be approximated by the union of one or more $\Sigma_j(t)$ at any time instance $t\geq 0$.
Finally we also introduce 
the variables $\cbra{\bar c_i}_{i=1}^K$ such that 
\begin{equation}
\bar c_j = i \Leftrightarrow j\in J_i.
\label{eq:hatc}
\end{equation}
Now the hybrid NDT model can be written in the form: 
\begin{equation}
\begin{cases}
    y(t) &= N_{\sigma}(t,x)\\
    \sigma(t) &= \sum_j \mathds{1}_{\sbra{x\in\Sigma_j(t)}} \bar c_j
\end{cases},\quad t\in \mathbb R_+.
    \label{eq:hybridNDT}
\end{equation}
A depiction of the model parameters of the NDT representation 
\eqref{eq:hybridNDT} is shown in Fig. \ref{fig:hybridNDT}.
In the next section, we will make use of this NDT representation to construct a 
data-driven learning algorithm to identify and adapt the hybrid NDT.

\section{Data-based Hybrid NDT Identification}
\label{Sec:ODA}

In this section we will use a prototype learning method to represent the 
partition $\cbra{\Sigma_i(t)}_{i=1}^K$, estimate the variables $\cbra{\bar c_i}_{i=1}^K$, and approximate the models $\cbra{N_i(t,x)}_{i=1}^s$ of 
the hybrid NDT model \eqref{eq:hybrid}.

\subsection{Prototype-based Learning Model}
\label{sSec:learning}

To efficiently represent $\cbra{\Sigma_i(t)}_{i=1}^K$, we make use of a polyhedral Voronoi partition produced by a set of variables $\cbra{\rho_i(t)}_{i=1}^K$, $\rho_i(t)\in W$, $\forall t\in\mathbb R_+$,  according to the rule: 
\begin{equation}
\Sigma_i(t) = \cbra{r\in W: i=\argmin_j d(r,\rho_j(t))},\quad t\in\mathbb R_+,     
\label{eq:Sigma-partition}
\end{equation}
for a given dissimilarity measure $d$.
For the Voronoi regions $\Sigma_i$
to be polyhedral, we select $d$ to be a Bregman divergence \citep{mavridis2025real}, 
a family of dissimilarity measures that includes, among others, the squared (weighted) Euclidean distance, and the Kullback-Leibler divergence
\citep{mavridis2023online,mavridis2023annealing}.
To minimize the updates and data consumption of the learning process, 
we assume that $\rho_i(t)$ only change during discrete events 
(e.g., cell overload or major network changes and workspace alterations) 
and model the dynamics of $\rho_i(t)$ as a first-order filter of the form:
\begin{equation}
\dot \rho_i(t) = - \gamma_\rho \rho_i(t) + \bar \rho_i^{(k)},\ \rho(0)=\rho_0,
\label{eq:rhodot}
\end{equation}
for all $i\in\cbra{1,\ldots,K}$, some $\gamma_\rho>0$, and $t\in\mathbb R_+$. The parameters $\tilde \rho_i^{(k)}$ are event-triggered functions
that change in discrete time instances and define the equilibrium of \eqref{eq:rhodot}.
In that way, the learning algorithm needs to estimate the optimal 
number and locations of the constant parameter vectors $\cbra{\bar \rho_i}_{i=1}^K$, as well as the association parameters $\cbra{\bar c_i}_{i=1}^K$, detect possible changes through an appropriate criterion, and provide a fast adaptation mechanism.

In addition to the above problem, which can be viewed as a combined clustering and classification problem, the estimation of 
$\cbra{N_i(t,x)}_{i=1}^s$ becomes an additional regression problem.
By introducing a set of constant models $\cbra{\bar Q_i(x)}_{i=1}^K$ as local models for each region $\Sigma_i$, we model the time dynamics of each $N_i(t,x)$ as:
\begin{equation}
\begin{aligned}
\pder{}{t} N_i(t,x) &= - \gamma_N N_i(t,x) + \sum_{j=1}^K \mathds{1}_{\sbra{x\in\Sigma_j}}  \bar Q_j^{(k)}(x) + \Delta_i^{(k)}(t),\\
N_i(0,x)&=N_{i}^0(x),\ i\in\cbra{1,\ldots,s},\ x\in W,\ t\in \mathbb R_+,
\end{aligned}
\label{eq:Ndot}
\end{equation}
for some $\gamma_N>0$. Model \eqref{eq:Ndot} is a first-order filter following the local models $\cbra{\bar Q_i^{(k)}(x)}_{i=1}^K$. 
The term $\Delta_i^{(k)}(t)$ models time-dependent changes that affect the
operation of cell $I_i$, such as cell overload or malfunction. 
It is an adaptation mechanism that can only be exploited through the 
use of a hybrid system of the form \eqref{eq:hybridNDT}.
These time-dependent terms will be discussed in Section \ref{Sec:TimeDependency}.

While there are many candidates for modeling $\bar Q_i(x)$, 
such as linear or quadratic functions, artificial neural networks, or Gaussian process regression models, we further compress the learning process by assuming constant models, i.e.,
\begin{equation}
\bar Q_i(x) = \bar Q_i,\quad \forall x\in W,\quad \forall i=1,\ldots,K.
\end{equation}
Since $\bar Q_i(x)$ are local models within the polyhedral regions $\Sigma_i(t)$, spatial information is already stored in the shape, position, and area of $\Sigma_i(t)$. 
The optimization problem is thus transformed from a model regression problem to a partition optimization such that local constant models provide a good approximation of the spatial characteristics of the NDT model.
As a result, the hybrid NDT model identification problem can be written as follows: 

\medskip
\begin{problem}[Hybrid NDT Identification]
Estimate the number $K$ and the parameters $\cbra{\bar \rho_i}_{i=1}^K$, $\cbra{\bar c_i}_{i=1}^K$, and $\cbra{\bar Q_i}_{i=1}^K$, such that the partition $\cbra{\Sigma_i(t)}_{i=1}^K$ minimizes both error terms $d\pbra{\sigma(t),c(t,x)}$ and $d\pbra{y(t),Q(t,x)}$ for a given 
Bregman divergence $d$, as described above.    
\label{prb:HybridNDT}
\end{problem}

Problem \ref{prb:HybridNDT} can be written as a combined clustering and classification problem of the form:
\begin{equation}
\begin{aligned}
&\minimize_{\cbra{K,\bar \rho_i,\bar Q_i, \bar c_i}} 
\E{\sum_{i=1}^K \mathds 1_{\sbra{r\in \bar\Sigma_i}}  d_c(Z, c, \mu_i, \bar c_i)}\\
\quad\quad
&\text{s.t.}\ \bar\Sigma_i = \cbra{r\in W: i=\argmin_j d(r,\bar\rho_j)},    
\end{aligned}
\label{eq:clustering-problem}
\end{equation}
where the random observation vector:
\begin{equation}
Z = \begin{bmatrix} x(t)\\ Q(t,x) \end{bmatrix}\in \bar W\subseteq R^{d+l}
\label{eq:X}
\end{equation}
is defined in a probability space $\pbra{\Omega, \mathcal{F}, \mathbb{P}}$, 
$c$ is a random variable following the distribution of $c(t,x)$, 
$\mu_i$ is the augmented codevector:
\begin{equation}
\mu_i := \begin{bmatrix} \bar \rho_i\\ \bar Q_i \end{bmatrix}\in \bar W,\ 
i=1,\ldots,K,    
\label{eq:mu-definition}
\end{equation}
and $d_c$ is a dissimilarity measure of the form:
\begin{equation}
d^c(x,c_x,y,c_y) = \begin{cases}
    d(x,y),~ c_x=c_y \\
    \infty,~ c_x\neq c_y			
\end{cases}
\label{eq:dc}
\end{equation}
motivated by the combined clustering and classification problem formulation in \citep{mavridis2023online,mavridis2023multi}.
%

Problem \eqref{eq:clustering-problem} is a hard clustering problem
with respect to the parameters $\cbra{\bar \rho_i,\bar Q_i, \bar c_i}_{i=1}^K$ with 
known $K$.
In Section \ref{sSec:oda}, we make use of the Online Deterministic Annealing (ODA) method 
as a homotopy optimization approach to progressively estimate the optimal parameters and 
the minimum number $K$ needed using an online, gradient-free recursion.

\subsection{Online Deterministic Annealing}
\label{sSec:oda}

The Online Deterministic Annealing method is a 
a recursive stochastic optimization algorithm that can solve problems of the form 
\eqref{eq:clustering-problem}, by progressively estimating the number $K$ of the 
augmented codevectors $\cbra{\mu_i}_{i=1}^K$
\citep{mavridis2023online,mavridis2023annealing}.

First 
a quantizer $Q:\bar W \rightarrow \bar W$ 
is defined as a discrete random variable in the same probability space 
with countably infinite domain $\mu := \cbra{\mu_i}$.
Then the following multi-objective optimization problem is formulated
\begin{equation}
\minimize_\mu F_\lambda(\mu) := (1-\lambda) D(\mu) - \lambda H(\mu),\ \lambda\in[0,1),
\label{eq:F}
\end{equation}
where the term
\begin{equation*}
\begin{aligned}
D(\mu) &:= \E{d\pbra{Z,Q}} 
=\int p(z) \sum_i p(\mu_i|z) d(z,\mu_i) ~\textrm{d}z
\end{aligned}
\label{eq:softVQ}
\end{equation*}	
is a generalization of the objective in (\ref{eq:clustering-problem}), and 
\begin{equation*}
\begin{aligned}
H(\mu) &: 
=H(Z) - \int p(z) \sum_i p(\mu_i|z) \log p(\mu_i|z) ~\textrm{d}z
\end{aligned}    
\end{equation*}
is the Shannon entropy.
This is now a problem of finding the locations $\cbra{\mu_i}$ and the 
corresponding probabilities
$\cbra{p(\mu_i|z)}:=\cbra{p(Q=\mu_i|Z=z)}$.
The Lagrange multiplier $\lambda\in[0,1)$ controls the trade-off between 
$D$ and $H$.
The entropy term, introduces several properties to the approach 
%
including robustness with respect to initial conditions 
\citep{mavridis2023online,mavridis2022risk}.
%
In addition, reducing the values of $\lambda$ defines 
a direction that resembles an annealing process
\citep{mavridis2023online} 
and a bifurcation phenomenon, according to which
the number $K$ of the codevectors increases as the value of
$\lambda$ decreases below certain data-dependent values
\citep{mavridis2023annealing,mavridis2023multi}. 

As shown in \citep{mavridis2023annealing}, a sufficient solution to \eqref{eq:F}, for a fixed value of $\lambda$, is given by
\begin{equation}
\mu_i^* = \E{Z|\mu_i} = \frac{\int z p(z) p^*(\mu_i|z) ~\textrm{d}z}{p^*(\mu_i)},
\label{eq:mu_star}
\end{equation}
where
\begin{equation}
p^*(\mu_i|z) = \frac{e^{-\frac{1-\lambda}{\lambda}d(z,\mu_i)}}
{\sum_j e^{-\frac{1-\lambda}{\lambda}d(z,\mu_j)}} ,~ \forall z\in S,
\label{eq:gibbs}
\end{equation}
as long as $d$ is a Bregman divergence.
In addition, 
the following Lemma 
constructs a gradient-free stochastic approximation algorithm 
that recursively estimates this solution:

\medskip
\begin{lemma}[\citep{mavridis2023online}]
The sequence $\mu_i(n)$ constructed by the recursive updates
\begin{equation}
\begin{cases}
\rho_i(t+1) &= \rho_i(t) + \beta(t) \sbra{ s_i \hat p(\mu_i|z_t) - \rho_i(t)} \\
\sigma_i(t+1) &= \sigma_i(t) + \beta(t) \sbra{ s_i z_t \hat p(\mu_i|z_t) - \sigma_i(t)}
\end{cases}
\label{eq:oda_learning1}
\end{equation}
where $z_t\sim Z$, 
$s_i:=\mathds{1}_{\sbra{\bar c_i=c}}$, 
$\sum_t \beta(t) = \infty$, $\sum_t \beta^2(t) < \infty$,
and the quantities $\hat p(\mu_i|z_t)$ and $\mu_i(t)$ 
are recursively updated 
as follows:
\begin{equation}
\begin{aligned}
\mu_i(t) = \frac{\sigma_i(t)}{\rho_i(t)},\quad
\hat p(\mu_i|z_t) = \frac{\rho_i(t) e^{-\frac{1-\lambda}{\lambda}d(z_t,\mu_i(t))}}
			{\sum_i \rho_i(t) e^{-\frac{1-\lambda}{\lambda}d(z_t,\mu_i(t))}}, 
\end{aligned}
\label{eq:oda_learning2}
\end{equation}
converges almost surely to a solution of 
(\ref{eq:mu_star}).
\label{lem:ODA}
\end{lemma}
%



Lemma \ref{lem:ODA} describes how to solve the optimization 
problem for a given value of the parameter $\lambda$.
In the online deterministic annealing approach,  
a sequence of optimization problems with decreasing values 
of $\lambda$, forming a homotopy optimization method.
%
%
It is shown that the unique values of the set $\cbra{\mu_i}$ 
that solves \eqref{eq:F},
form a finite set of $K(\lambda)$ values, referred to as ``effective codevectors''.
%
At high values ($\lambda\rightarrow 1$),~(\ref{eq:gibbs}) yields
uniform association probabilities $p(\mu_i|z)=p(\mu_j|z),\ \forall i,j, \forall z$, 
and as a result of (\ref{eq:mu_star}), all pseudo-inputs are located at the same point
$\mu_i = \E{Z},\ \forall i $
which means that there is one unique ``effective'' codevector given by $\E{X}$.
As $\lambda$ is lowered below a critical value, a bifurcation phenomenon occurs, 
when the number of ``effective'' codevectors increases, 
which describes an annealing process \citep{mavridis2023online,mavridis2023multi}.
Mathematically, this occurs when the existing solution $\mu^*$ given by (\ref{eq:mu_star}) 
is no longer the minimum of the free energy $F^*$,
as the temperature $\lambda$ crosses a critical value.
%
%

In other words, the number of codevectors increases countably many times as 
the value of $\lambda$ decreases, and an algorithmic implementation needs only
as many codevectors in memory as the number of ``effective'' codevectors.
In practice.
the bifurcation points are detected
by introducing perturbing pairs of effective codevectors
at each temperature level $\lambda$. 
In the case of classification, a perturbed copy of each effective codevector is produced for each unique known class label 
$\underline{c}_j \in \underline{c}$ as follows:
\begin{equation}
\begin{aligned}
\bar \mu &\gets \bigcup_{\bar \mu_i\in \bar \mu} \cbra{\bar\mu_i+\delta_j}_{j\in|\underline c|} \\
\bar c &\gets \bigcup_{i\in |\bar \mu|} \cbra{\bar c_j}_{c_j\in \underline c}.
\end{aligned}
\label{eq:perturbation}
\end{equation}
%
The newly inserted codevectors will merge with their pair if 
a critical temperature has not been reached and separate otherwise, 
reflecting a genetic-type approach
\citep{mavridis2023annealing}.
%
%
A detailed discussion on the implementation of the 
original online deterministic annealing algorithm,
its complexity,
and the effect of its parameters,
can be found in \citep{mavridis2023online,mavridis2023annealing,mavridis2023multi}.

\section{Time Dependency}
\label{Sec:TimeDependency}

The learning algorithm proposed in Section \ref{Sec:ODA} 
estimates the parameters
$\{\bar \mu_j,\bar Q_j, \bar c_j \}_{j=1}^K$, along with the number $K$.
These parameters are estimated in discrete time, according to the frequency and availability of the observations.
Equations \eqref{eq:rhodot} and \eqref{eq:Ndot} constitute filtering equations for a stable continuous-time approximation of the 
hybrid NDT. 
In this section, we focus on the design of event-trigger mechanisms to detect changes in the values $\{\bar \mu_j,\bar Q_j, \bar c_j \}_{j=1}^K$, which, by construction, represent local averages in the region $\bar \Sigma_i$. Therefore, a change in these local average values signifies a time-dependent drift in the model, which triggers a new training cycle and/or corrections terms, such as the quantities $\Delta_i^{(k)}(t)$ in \eqref{eq:Ndot}.

\subsection{Regression Error Event Trigger}

We detect a communication quality regression error event by the condition:
\begin{equation}
    \|Q(t,x) - \sum_{j=1}^K \mathds{1}_{\sbra{x\in\Sigma_j}}  \bar Q_j \| \geq \epsilon_q
    \label{eq:regression-trigger}
\end{equation}
for some $\epsilon_q>0$.
This type of drift can signify local or global network changes.
Therefore, in addition to the parameters $\cbra{\bar Q_i}$, 
the partition $\cbra{\bar \Sigma_i}$ needs to be updated as well.
The drift detection triggers the algorithm to increase the parameter $\lambda$ by a predefined amount, and continue the stochastic approximation updates of the learning algorithm. Increasing the parameter $\lambda$ naturally resets the algorithm to focus on larger changes. The parameter $\lambda$ decreases back to its original value according to the algorithm, having adjusted to the changes without the need to learn the model from scratch.

\subsection{Classification Error Event Trigger}

We detect a cell ID mode classification error event by the condition:
\begin{equation}
   \sum_i \|c(t,x_i) - \sum_{j=1}^K \mathds{1}_{\sbra{x_i\in\Sigma_j}} \bar c_j\|\geq \epsilon_s,
   \label{eq:classification-trigger}
\end{equation}
for all $x_i$ in a rolling window of observations and some $\epsilon_s>0$.
Similar to the regression error, the misclassification detection triggers the algorithm to increase the parameter $\lambda$ by a predefined amount, and continue the stochastic approximation updates of the learning algorithm,
avoiding the need to learn the model from scratch.

\subsection{Cell-Specific Error Event Trigger}

In a 6G network with multiple base stations, it is often the case that 
the communication quality between a base station $I_i$, $i\in\cbra{1,\ldots,m}$, and the UEs connected to them is affected due to a cell malfunction or high user demand. 
In this case, the user experiences high latency that is often reflected in the  the SINR metric. A condition can be constructed 
if a latency or SINR component $Q_\tau(t,x)$, $\tau\in\cbra{1,\ldots,l}$, is observed by the network, such that:
\begin{equation}
    \|Q_\tau(t,x) - \sum_{j=1}^K \mathds{1}_{\sbra{x\in\Sigma_j}} \bar Q_{j_\tau} \| \geq \epsilon_\tau
    \label{eq:cell-trigger}
\end{equation}
for some $\epsilon_\tau>0$.
This induces a drift in the hybrid NDT model, that is shared among the mode $\sigma(t)=i$, and is usually temporary. In that case, the correction term 
$\Delta_i^{(k)}(t)$ in \eqref{eq:Ndot} is used, defined as follows:
\begin{equation}
\Delta_i^{(k)}(t) = 
\begin{cases}
    Q(t,x) - \sum_{j=1}^K \mathds{1}_{\sbra{x\in\Sigma_j}}  \bar Q_j, 
    \quad t_k<t<T \\ 
    0,\quad t>t_k+T
\end{cases}
\label{eq:delta}
\end{equation}
where $t_k$ represents the time of the event trigger ($k$-th discrete timestep), and $T>0$ is a predefined time window.
The term $\Delta_i^{(k)}(t)$ fades after time $T$, and no re-training is necessary.

\section{Experimental Results}
\label{Sec:Results}

We illustrate the properties and evaluate the performance 
of the proposed hybrid NDT using data observations acquired by a single UE operating in a closed workspace over a real 5G network.
%
%

\subsection{Experimental Apparatus}

The data used to demonstrate the proposed method was collected in a real 5G testbed \citep{hernandez2025end}, located at Ericsson's headquarters in Sweden. The testbed embraces the concept of device-edge-cloud continuum, allowing for the seamless integration of compute platforms and devices, thanks to the use of cellular network. The testbed is equipped with monitoring and exposure tools that allow for the extraction of network metrics from user equipment, base station, and core network.

The datasets we use in this work was collected autonomously by a mobile robot in an outdoor area covered by two 5G base stations. The set of measurements $Q_i$ that were selected are time, position, user RSRP, and user SINR with an average frequency of $20Hz$.

To validate our methodology, we test two different scenarios against a neural network-based NDT trained as a regression problem with stochastic gradient descent on a fully connected neural network of $n_{n}=100$ neurons with ReLu activation function.

\begin{figure}[ht]
\centering
\includegraphics[trim=30 30 10 20,clip,width=\linewidth]{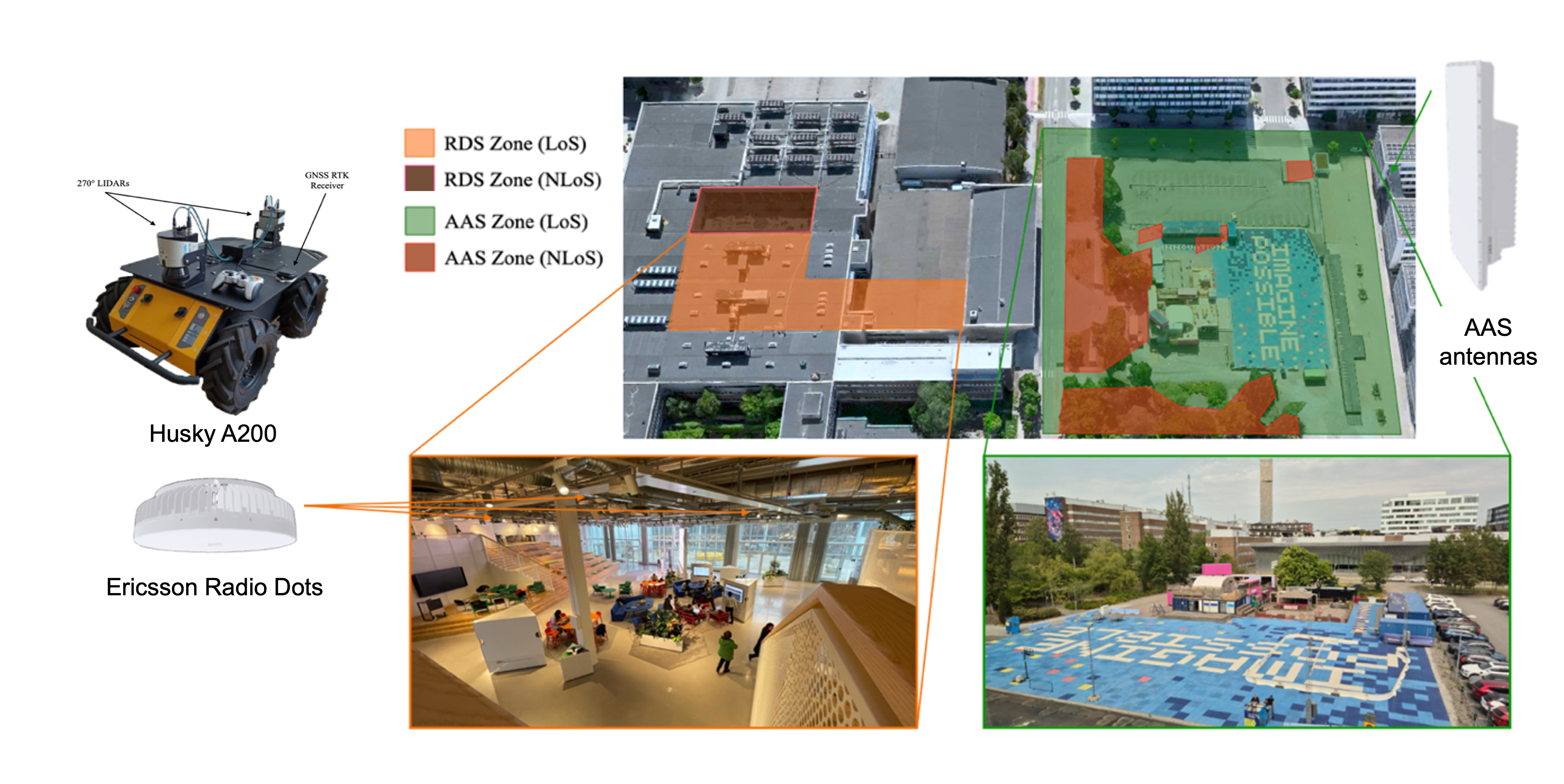}
\caption{Experimental apparatus at Ericsson's headquarters, Stockholm, Sweden.}
\label{fig:husky}
\end{figure}
%

\subsection{Hybrid NDT Approximation with Model Drift}

In the first scenario, the mobile robot scans the workspace in a predefined trajectory. The hybrid NDT converges after approximately $20$ seconds having collected approximately $500$ observations 
and having produced $K=38$ codevectors $\bar\rho_i$ (respectively regions $\bar \Sigma_i$).
A regression error event-trigger is identified at time $T_m\approx 30$ s
using \eqref{eq:regression-trigger}, and the hybrid NDT raises the parameter $\lambda$ by a factor of $10\%$, and continues the recursion 
in \eqref{eq:oda_learning1},\eqref{eq:oda_learning2} using newly collected data.
At the same time, the neural network model continues a gradient descent recursion using new data. 

\begin{figure}[ht]
\centering
\includegraphics[trim=0 0 -20 0,clip,width=0.49\textwidth]{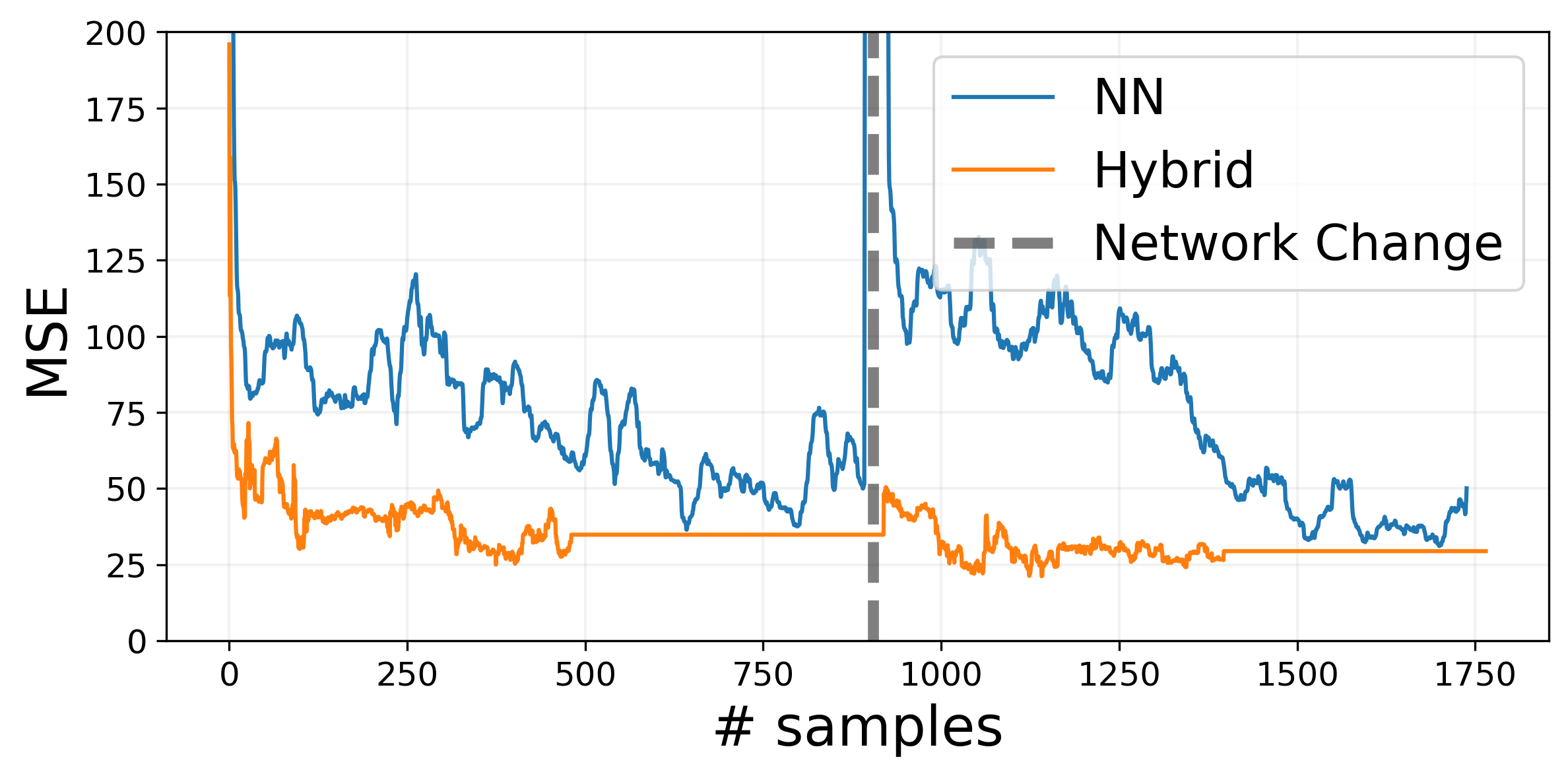}
\caption{Comparison of the training behavior of the proposed hybrid NDT with a neural network regressor.}
\label{fig:training}
\end{figure}

\begin{figure}[ht]
\centering
\includegraphics[trim=0 0 0 0,clip,width=0.45\textwidth]{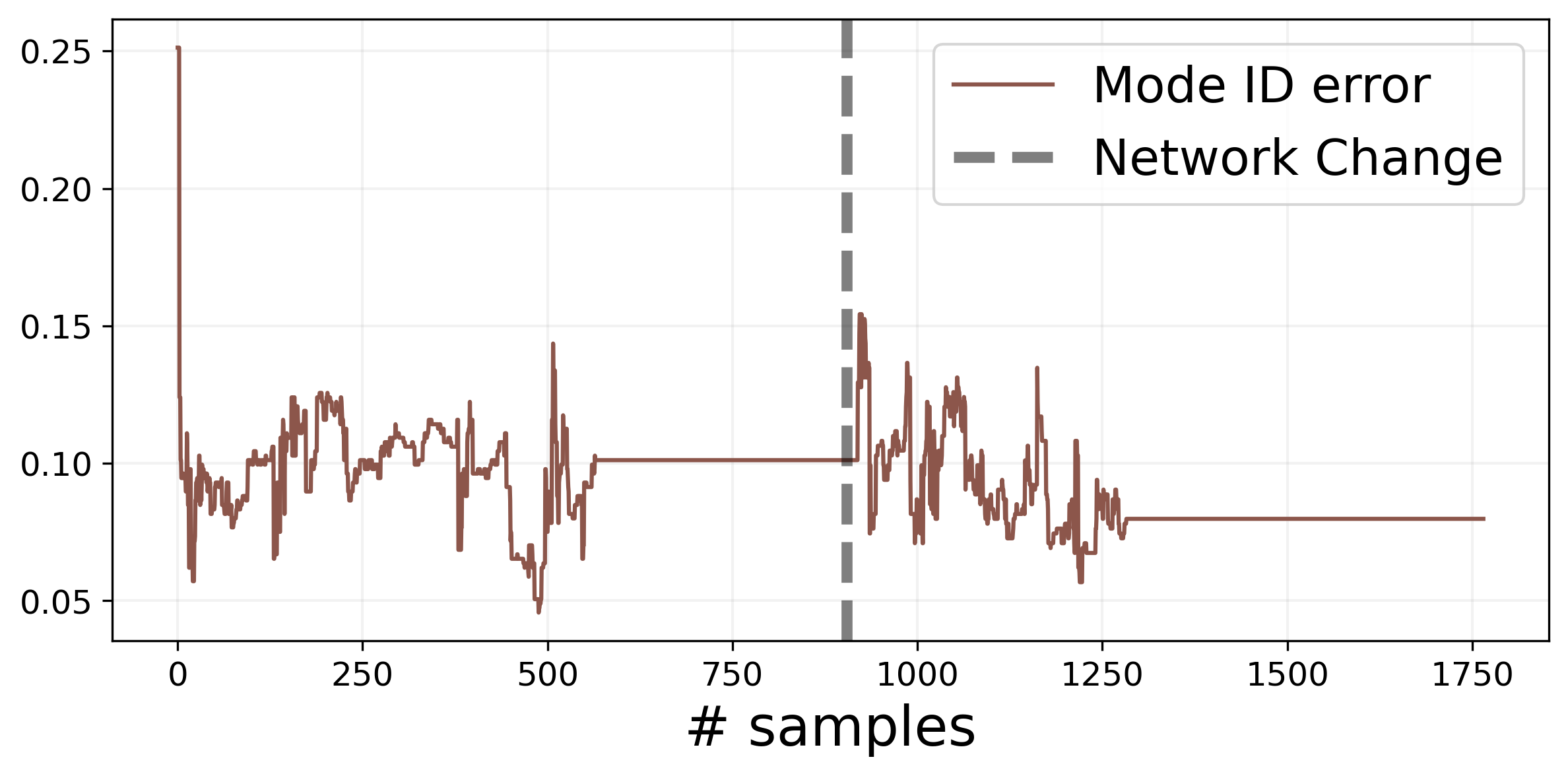}
\caption{Mode identification error of the hybrid NDT.}
\label{fig:classification-training}
\end{figure}

\begin{figure}[ht]
\centering
\begin{subfigure}[b]{0.48\textwidth}
\centering
\includegraphics[trim=9 0 75 0,clip,width=0.45\textwidth]{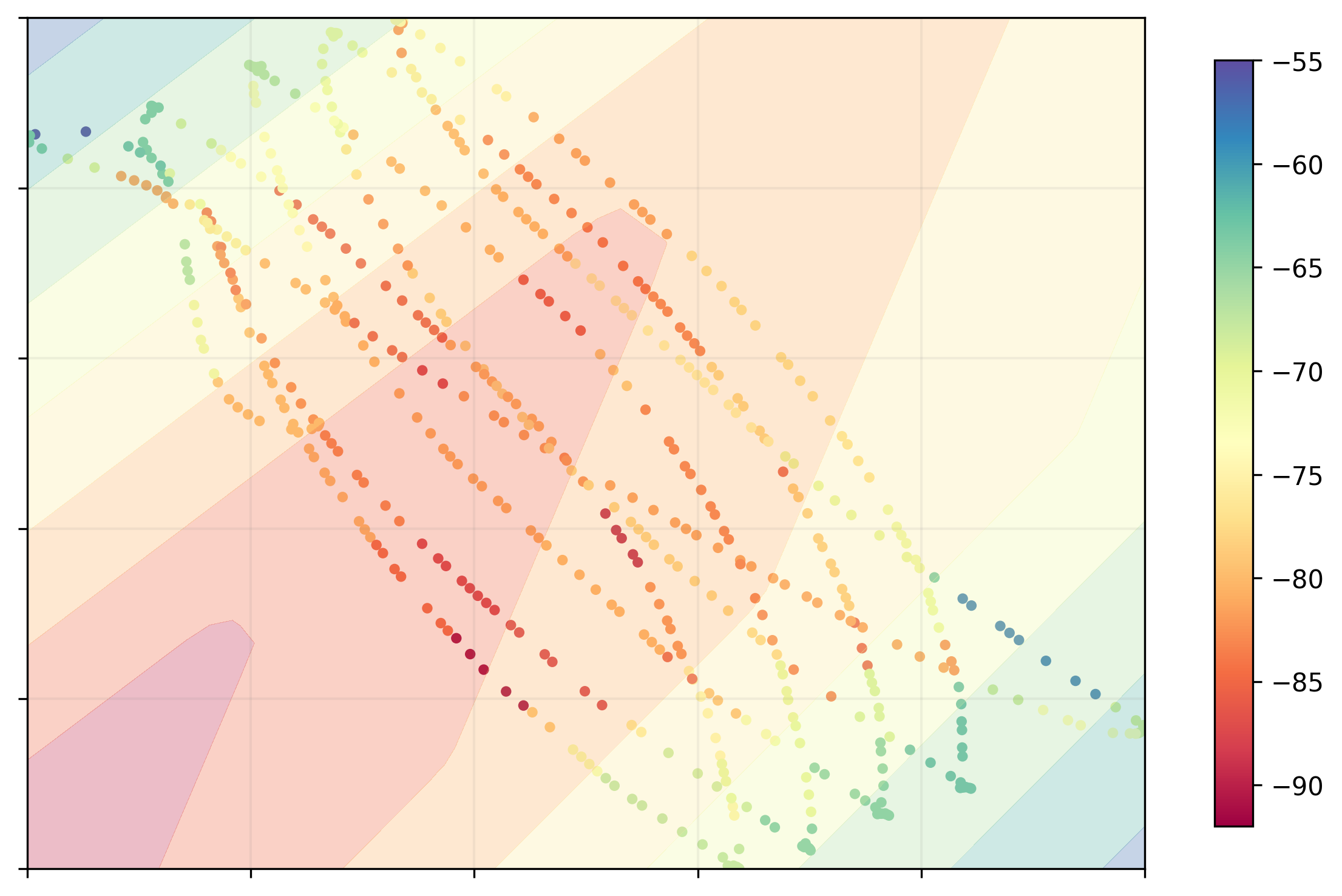}
\includegraphics[trim=9 0 75 0,clip,width=0.45\textwidth]{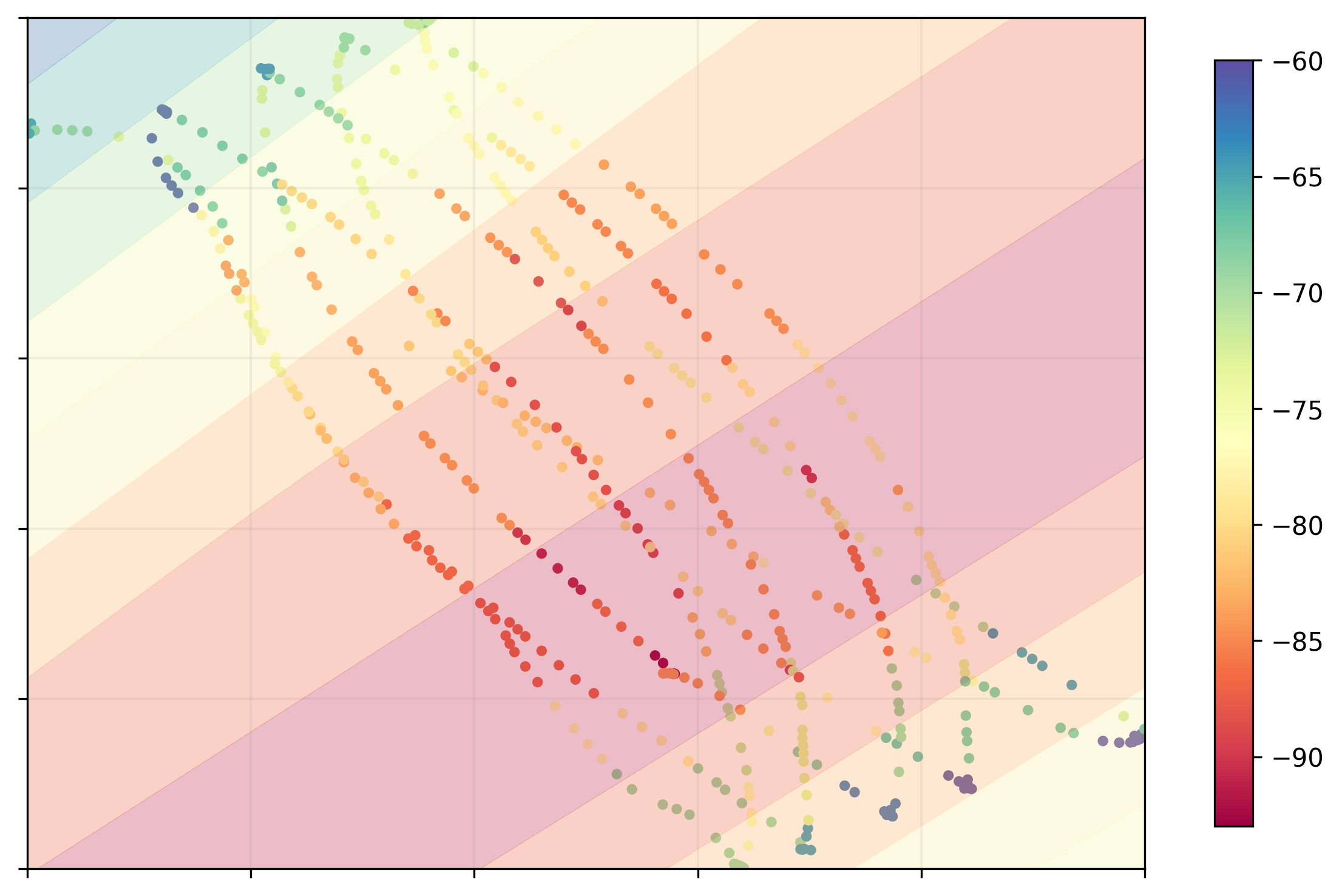}
\includegraphics[trim=480 22 0 0,clip,width=0.058\textwidth]{nn-rsrp-2}
\caption{Neural network approximation.}
\end{subfigure}
\begin{subfigure}[b]{0.48\textwidth}
\centering
\includegraphics[trim=9 0 75 0,clip,width=0.45\textwidth]{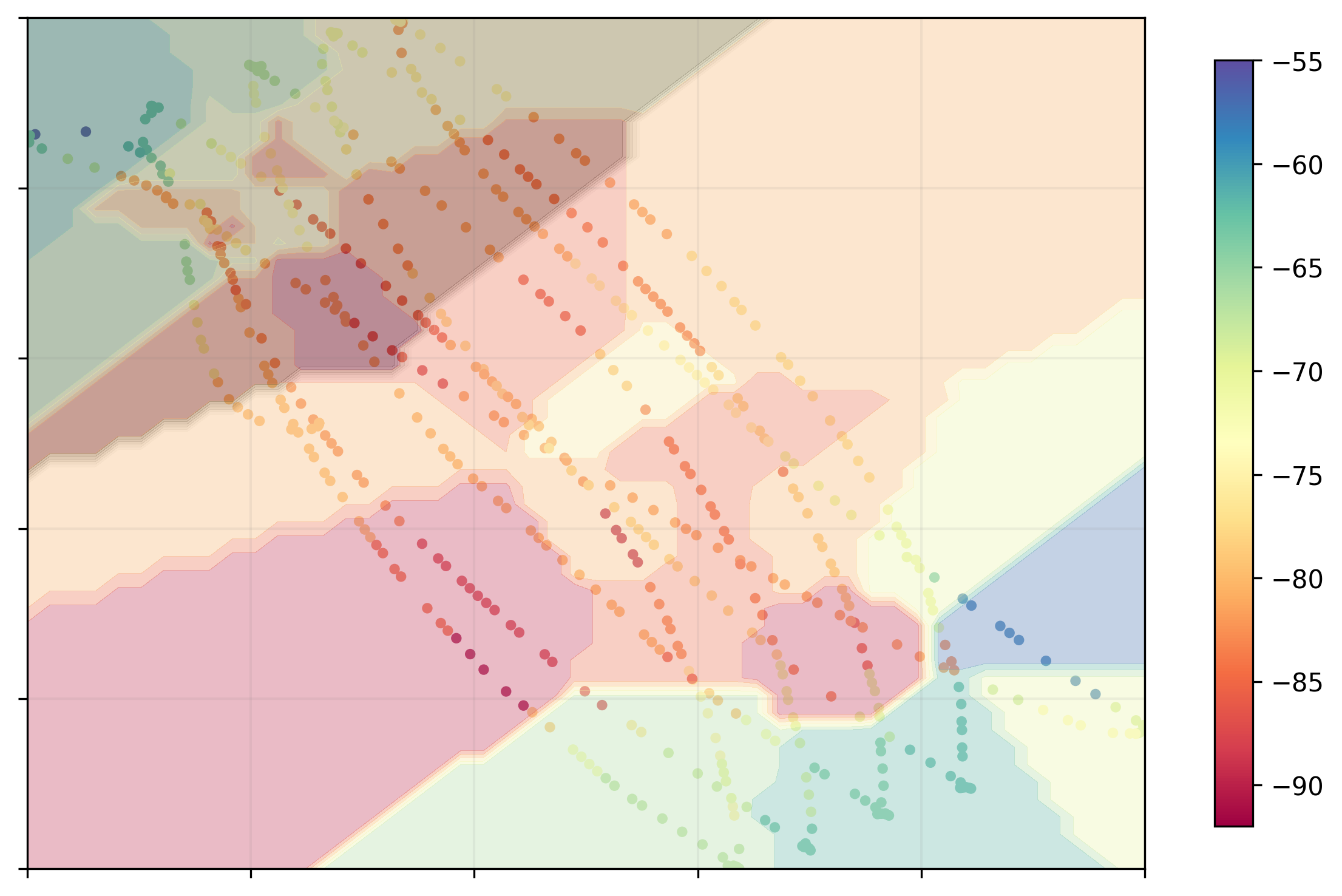}
\includegraphics[trim=9 0 75 0,clip,width=0.45\textwidth]{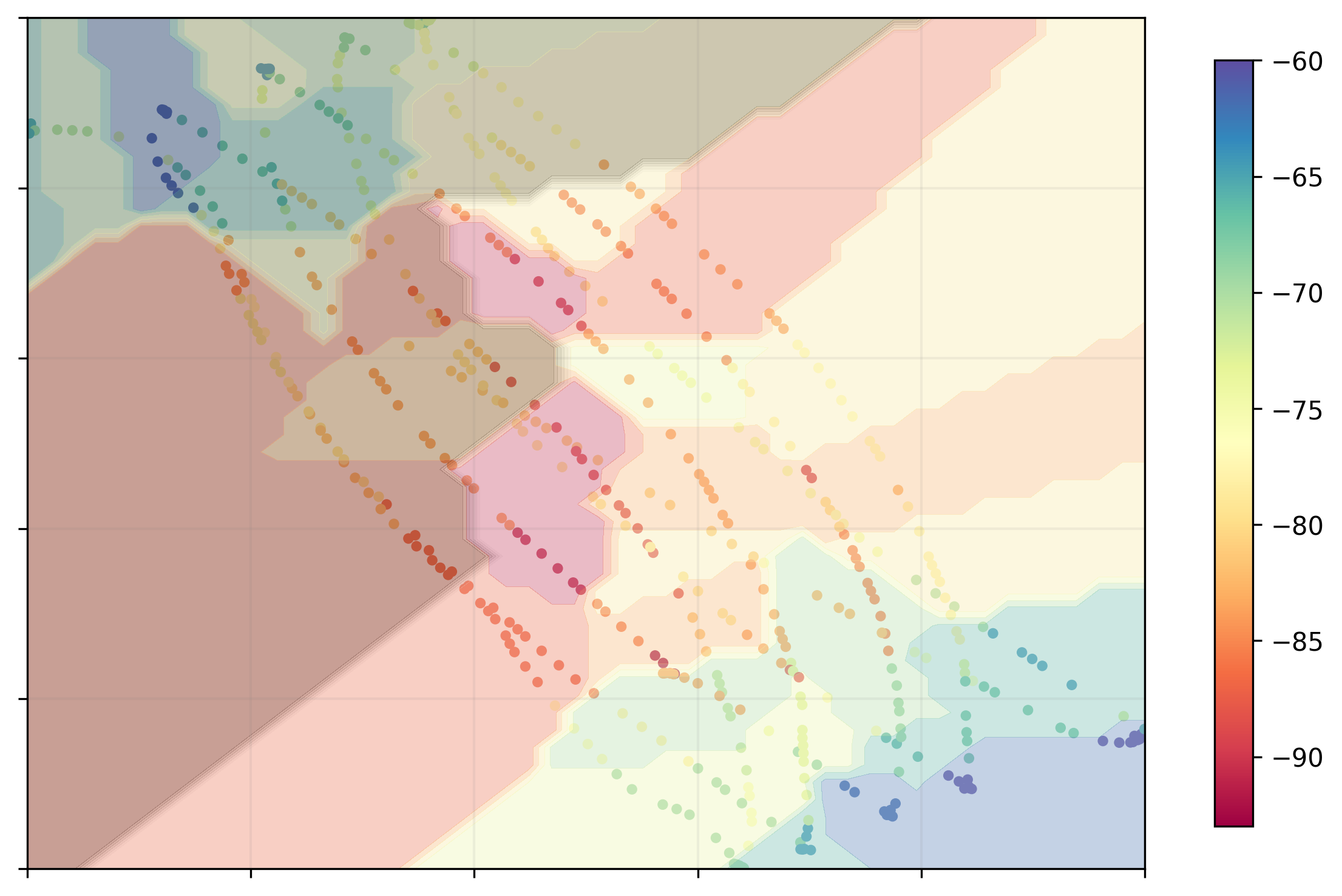}
\includegraphics[trim=480 22 0 0,clip,width=0.058\textwidth]{oda-rsrp-2}
\caption{Hybrid NDT approximation.}
\end{subfigure}
\caption{RSRP approximation before (left) and after the network change (right). Shaded areas represent the estimation of mode $S_1$. Data samples are also depicted.}
\label{fig:NDTaproximation}
\end{figure}

In Fig. \ref{fig:training}, we illustrate the training behavior of the two approaches. Notice how the proposed hybrid NDT method is faster to converge to low mean-square-error (MSE) values, faster to converge (lower number of observations used), and faster to adapt to the network change.
In addition, the classification error that corresponds to the identification of the modes $S_i$, $i=1,2$, is depicted in Fig. \ref{fig:classification-training}.
Finally, in Fig. \ref{fig:NDTaproximation} we depict the hybrid NDT and the neural network-based RSRP approximation before and after the network change. 
Notice how the lack of structure in black-box regression models can produce biased representations that, even though they present low regression error, can not serve as a basis for fast adaptation to local changes.

%

\subsection{Hybrid NDT Adaptation on Cell Malfunction}

\begin{figure}[ht]
\centering
\includegraphics[trim=0 0 0 0,clip,width=0.48\textwidth]{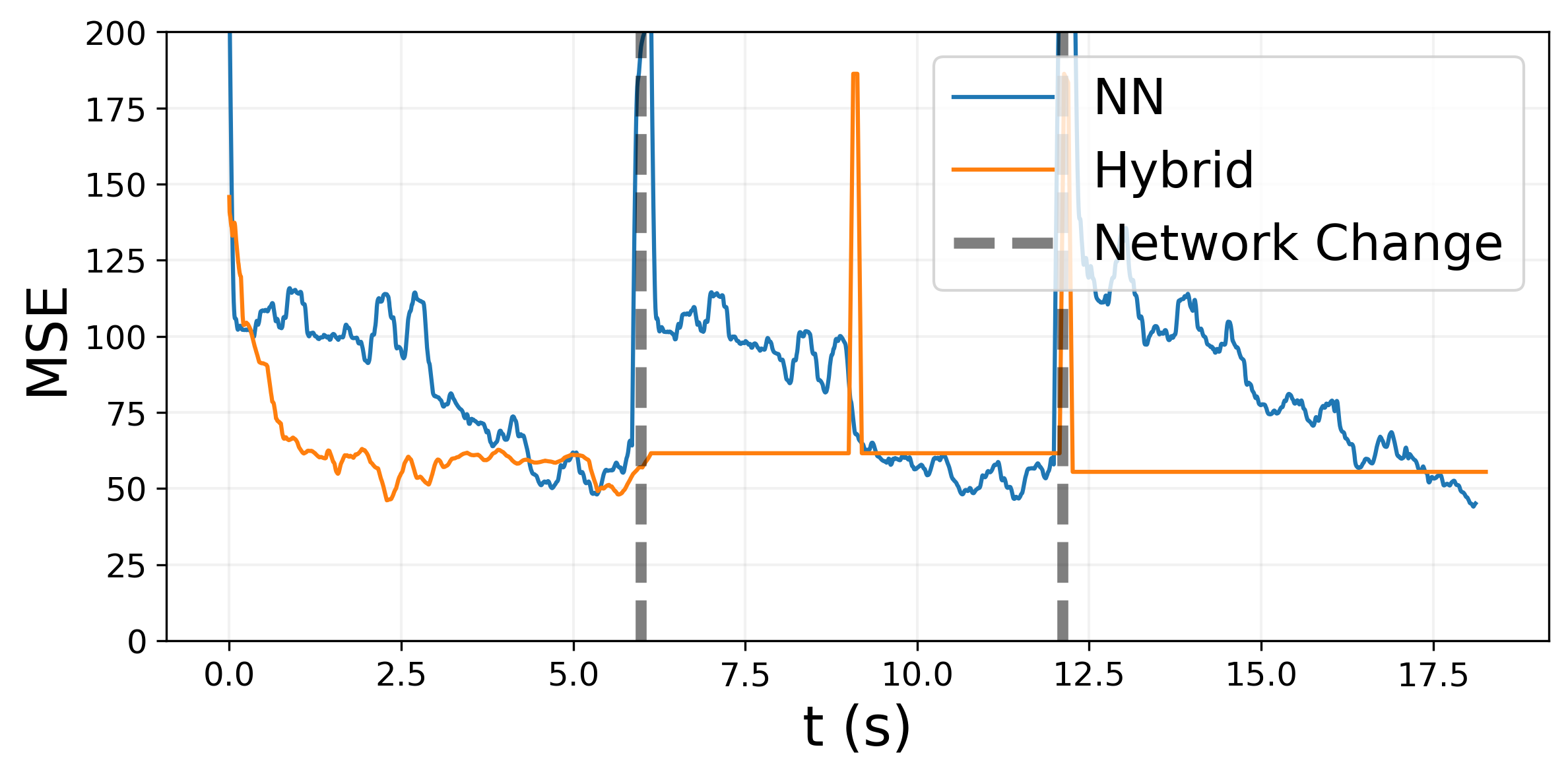}
\caption{SINR approximation error during cell malfunction. The hybrid NDT model is not retrained. Instead the term $\Delta_i^{(k)}$ in \eqref{eq:Ndot} is used. The spikes correspond to the end of the predefined time duration $T$ in \eqref{eq:delta}.}
\label{fig:delta}
\end{figure}

In the second scenario, we simulate a temporary cell malfunction by cutting down the user SINR measurements for users connected to cell $I_1$ to zero for approximately $T_0=6$ seconds. 
In Fig. \ref{fig:delta} we show the training behavior of the proposed hybrid NDT against a neural network-based NDT in estimating the SINR component. 

Notice that the hybrid NDT is only training within the first $6$ seconds. 
When the cell malfunction takes place, the $\Delta_i^{(k)}$ term in \eqref{eq:Ndot}, \eqref{eq:delta} is used. 
In particular, all $\bar Q_j$ terms associated with the regions $\bar\Sigma_j$ that make up $S_1$ (i.e., $\bar c_j=1$) are canceled in \eqref{eq:Ndot}, and $N_1(t,x)$ converges to zero as the solution of a stable linear time-invariant differential equation (with respect to time). 
In this experiment, a time window duration $T=3$s was chosen.
For this reason, a spike appears temporarily in the training error at $t\approx 9$s.
At that point, a cell-specific error event-trigger of the form \eqref{eq:cell-trigger} 
gets activated again.  

After the cell malfunction period is over, the model $N_1(t,x)$ converges back to its initially leaned $\bar Q_j$ values.
That is to say, no more training takes place after $t\approx 6$s.
This is in contrast to the neural network-based NDT model that needs retraining from scratch 
after any observed change in the network, as shown in Fig. \ref{fig:delta}.

%

\section{Conclusion and Future Work}
\label{Sec:Conclusion}

In this work, we focused on the representation of NDTs that model the communication quality properties of a multi-cell, dynamically changing wireless network over a workspace populated with multiple, moving UEs.
We explored the advantages of modeling the NDT as a hybrid system, 
where the discrete modes that are associated with the different base stations, as well as areas of the workspace with similar network characterisitics.
We further proposed an annealing optimization-based learning algorithm, driven by online data measurements collected by the UEs, 
to identify and continuously improve the hybrid NDT.
Finally, simulations with real experimental data showed that the proposed hybrid NDT can yield higher memory and computational efficiency, lower data consumption, and increased adaptation speed with respect to network changes.

Ongoing and future research endeavors are focused on the fusion of the proposed hybrid NDT identification 
methodology in motion planning applications, where the explainability, adaptability, data efficiency, and compressed information rate properties could lead to a combined communication-aware motion planning approach.


\ifx\ieee\undefined

\bibliography{bib_camp.bib,bib_learning.bib,bib_mavridis.bib,bib-switched.bib,bib_ndt.bib} 

\else

\balance
\bibliographystyle{IEEEtran} %
\bibliography{bib_camp.bib,bib_learning.bib,bib_mavridis.bib,bib-switched.bib,bib_ndt.bib} 

\fi

\end{document}